\documentstyle[12pt]{article}
\def\be{\begin{equation}}
\def\ee{\end{equation}}
\def\beqn{\begin{eqnarray}}
\def\eeqn{\end{eqnarray}}
\def\noin{\noindent}
\begin{document}
\begin{titlepage}
\title{Loop Equations for the d-dimensional\\
n-Hermitian Matrix model}
\author{Jorge Alfaro\\
Facultad de F\'\i sica\\Universidad Cat\'olica de
Chile\\Casilla 306, Santiago 22, Chile
\\jalfaro@lascar.puc.cl}
\date{November 8 1993}
\maketitle
\begin{abstract}

We derive the loop equations for the d-dimensional
n-Hermitian matrix model. These  are a consequence of
the Schwinger-Dyson equations of the model.
Moreover we show that in leading order of large $N$ the
loop equations form a closed set. In particular we derive
the loop equations for the $n=2^k$ matrix model recently proposed
to
describe the coupling of Two-dimensional
quantum gravity to conformal matter with $c> 1$.

\end{abstract}
\vfill
\begin{flushleft}
PUC-FIS 28-93 \\
November 1993
\end{flushleft}
\end{titlepage}
\newpage

\section{Introduction}

The last three years have seen a rapid development in the knowledge
relative to two-dimensional quantum gravity coupled to conformal matter
with $c\le 1$. This has been possible by the use of matrix model techniques
and the large $N$ expansion, combined with the double scaling limit
\cite{moore}
Very recently a new interpretation of the double scaling limit as a finite
size rescaling has emerged \cite{poul}\cite{hikami} based on the
renormalization
group applied to large $N$ models \cite{brezin}\cite{aldamg}.

Although the new techniques have provided a very deep understanding of
two-dimensional
quantum gravity coupled to $c\le 1$ conformal matter, a barrier has emerged
at $c=1$, avoiding the continuation of the previous results beyond this point.
The same barrier appears in the continuum treatment of non-critical strings.
\cite{polyakov}

Various approaches to surpass this problem have been suggested. Among them
we mention the use of reduced models to discuss the d-dimensional matrix model
\cite{luis}
which is the right model to discuss the d-dimensional string, Very recently
we have obtained the loop equations for this model for arbitrary $N$ and show
that they form a closed set in leading order of large $N$\cite{alfaro2}.
 For the moment
they have not been solved but we see them as an important exact result that
may permit to cross the barrier at $c=1$.

Last year a  $2^n$-Hermitian matrix model in zero dimensions have been
proposed to
describe a system with $c=n/2$ coupled to 2d gravity\cite{hikami2}.
So a solution of this
model is another way to cross the barrier.

Four years ago we obtained the loop equations for the zero dimensional
two-matrix model\cite{retamal}, extending the solution of Metha for the
partition function\cite{metha}. Last year we were able to solve these
equations
completely\cite{alfaro1}. See also \cite{staudacher}.

Motivated by these recent progress in the solution of the Two matrix model,
in the present work we will derive the
loop equations for the n-Hermitian matrix
model in any space-time dimension $d$ and
for arbitrary $N$ and show that they form
a closed set in leading
order of large $N$. They are a consequence of the Schwinger-Dyson equations
of the model and the
type of coupling among the different matrices. The couplings
we consider include the $q$-state Potts matter fields coupled
to Two Dimensional Gravity of \cite{hikami2} as a particular case.

This paper is organized as follows: In section 2 we review the Hidden BRST
method\cite{aldamg2} and use it to derive the loop equations in section 3.
In section 4 we discuss
in more detail the zero dimensional case. In particular,
we compare our equations with the two matrix equations of\cite{retamal}.
 In section 5 we present our conclusions and comment on future work.

\section{Hidden BRST Symmetry and Schwinger-Dyson Equations}

The usual method to derive Schwin\-ger-Dyson equations (SDe) involves
using the invariance of
the path integral measure under field translations.  Invariances of
the action itself are not relevant for this derivation.  In this
section we review an alternative derivation which has been
proposed some years ago \cite{aldamg2}: By making use of a BRST
symmetric extension of
any action $S$, we can derive all SDe's as BRST supersymmetric Ward
identities of the new action.
\vskip 1pc
To simplify the presentation, consider the path integral describing
one bosonic field
$\varphi(x)$:

\be
Z = \int[d\varphi] e^{-S(\varphi)} .
\ee
\noin
The functional measure $[d\varphi]$ in the last equation is invariant
under field translations:
\be
\varphi(x)\rightarrow\varphi(x) + \epsilon(x) .
\ee
\noin
The invariance of the functional measure implies the following
statement:
\be
\int[d\varphi]\frac{\delta[Fe^{-S(\varphi)}]}{\delta\varphi(x)} = 0
\ee
\noin
{}From (3) we get
\be
\langle\frac{\delta F}{\delta\varphi(x)}-F\frac{\delta S}
{\delta\varphi(x)}\rangle =
 0
\ee
\noin
for any $F=F(\varphi)$.  These are the SDe's
for the theory.  Notice
that, in general, $S$ is not invariant under (2) and that this lack
of invariance plays no role in the derivation of the SDe's  We now
introduce auxiliary fermionic variables $\psi(x)$, $\bar \psi(x)$ and insert
a trivial factor of unity
$$
1 = \int [d\bar\psi][d\psi] e^{-\int dx\bar\psi(x)\psi(x)}
$$
\noindent
inside the partition function.  The resulting action $\bar S$ is
invariant under the following transformation:
 \beqn
\bar S [\varphi,\psi,\bar\psi] &= S[\varphi] + \int dx \bar
\psi(x)\psi(x)\\
\delta \varphi(x) &= \psi (x) \\
\delta \psi (x) &= 0\\
\delta \bar \psi(x) &= -\delta S/\delta\varphi(x)
\eeqn

\noin
Associated with the BRST-like symmetry, (6)--(8) is a set of Ward
identities; for example, unbroken BRST implies:

$$
\langle\delta [F(\varphi) \bar\psi(y)]\rangle = 0,
$$
\noin
which is
$$
\langle\int dx \frac{\delta F}{\delta \varphi(x)}\psi(x)
    \bar\psi(y)
    -F \frac{\delta S}{\delta \varphi} \rangle=0 .
$$

\noin
Computing the average with respect to the fermionic variables we recover
eq. (4).  Notice that the BRST transformation we have just
defined commutes with any symmetry the original action $S$ may have.
In particular, if the original action
is invariant under a $U(N)$ group, the BRST
symmetry will commute with this group also and then its implications will be
true order by order in the $1/N$ expansion.

In the next section we use  the Schwinger-Dyson BRST  symmetry  to
derive the loop equations for the $d$-dimensional n-Hermitian matrix
model.

\section{Loop Equations}

In this section we will derive the loop equations for the model defined
by the action:
\be
S= \int d^dx[\sum_{k,l}tr(\frac{1}{2}d_{kl}\partial_\mu M(x,k)
\partial_\mu M(x,l)+
V_l(M(x,l))-\frac{1}{2}\sum_{i\ne j} c_{ij} tr M(x,i)M(x,j)]
\ee
$M(x,l)$ is an $N\times N$
Hermitian matrix defined on the (euclidean) space-time
point $x$, and $V_l$ is a local function of $M(x,l)$.

This action includes the sytems studied in \cite{hikami2}(we follow their
notation) if we take:
\beqn
d_{\tau\tau'}={1 \over {(2 \sinh 2\beta )^{n}}}\hat \tau \hat \tau '
\exp\beta\tau\cdot\tau '\\
\tau \cdot\tau '=\sigma_{1}\sigma '_{1}+\sigma_{2}\sigma '_{2}\dots
+\sigma_{n}\sigma '_{n}\\
\hat \tau =\sigma_{1}\sigma_{2}\dots\sigma_{n}\\
c_{\tau\tau'}=d_{\tau\tau'}
\eeqn
where $\sigma =1,2,\dots ,q$, for the q-state Potts matter fields coupled
to gravity.

The extended action and the BRST transformation are:
\beqn
S_{ext}&=S+\int d^dx\sum_l tr\bar\psi(x,l)\psi(x,l)\\
\delta M(x,l)&=\psi(x,l)\\
\delta\psi(x,l)&=0\\
\delta\bar\psi(x,l)&=-\frac{\delta S}{\delta M(x,l)}
\eeqn

Introduce the following representation of $M(x,l)$:
\be
M(x,l)=\sum_{\alpha=1}^N m_\alpha(x,l)T^\alpha(x,l),
\ee
where the projectors of the matrix $M(x,l)$ satisfy:
\beqn
\sum_{\alpha=1}^N T^\alpha(x,l)&=1\\
tr T^\alpha(x,l) T^\beta(x,l)&=\delta_{\alpha\beta}\\
T^\alpha(x,l) T^\beta(x,l)&=\delta_{\alpha\beta} T^\alpha(x,l) .
\eeqn
If $\delta$ denotes the BRST variation (see last section), we get the
following
BRST transformation for the projectors $T^\alpha$ and eigenvalues
$m_\alpha$
(see the Appendix of \cite{alfaro2}):
\beqn
\delta T^\alpha(x,l) &=
\sum_{\beta \neq \alpha}\frac{T^{\alpha}(x,l) \psi(x,l)
T^{\beta}(x,l) + T^{\beta}(x,l)\psi(x,l) T^{\alpha}(x,l)}
{m_{\alpha}(x,l)-m_{\beta}(x,l)} \\
\delta m_\alpha (x,l) &= {\rm tr}\psi(x,l) T^\alpha(x,l) .
\eeqn

Consider the following functional of $\lambda(x,l)$
\be
K_{ij}=\int d^dx\sum_l
\sum_{\alpha=1}^N\lambda_\alpha(x,l) T^\alpha(x,l)_{ij} .
\ee

The basic object we will use to write the loop equations is:
\be
u[\lambda]=e^{iK}
\ee
It fullfils the identity:
\be
\frac{\delta u[\lambda]}{\delta\lambda_\alpha(x,l)}=
\int_0^1 dt e^{itK} iT^\alpha(x,l) e^{i(1-t)K} ,
\ee
which implies:
\be
\frac{\delta {\rm tr} u[\lambda]}{\delta\lambda_\alpha(x,l)}=
i tr T^\alpha(x,l) u[\lambda]
\ee
The trace is taken with respect to the internal indices only.
{}From $u[\lambda]$ we can compute all symmetric combinations of products
of $T^\alpha$. For instance:
\be
\frac{\delta^2u[\lambda]}{\delta\lambda_{\beta_1}(x_1,l_1)
\delta\lambda_{\beta_2}(x_2,l_2)}|_{\lambda=0}=-\frac{
T^{\beta_2}(x_2,l_2)T^{\beta_1}(x_1,l_1)+
T^{\beta_1}(x_1,l_1)T^{\beta_2}(x_2,l_2)}
{2}
\ee
The Schwinger-Dyson equation is:
\be
\langle\delta tr [u[\lambda]\bar\psi(x,l)]\rangle=0
\ee
i.e.
\be
\langle tr [\delta u[\lambda]\bar\psi(x,l)+
tr [ u[\lambda]\delta\bar\psi(x,l)]\rangle=0 .
\ee
Since
\be
\delta\bar\psi(x,l)=
\sum_kd_{lk}\Box M(x,k)+ \sum_{j\ne l}c_{lj}M(x,j)-\Delta_l(M)
\ee
where
\be
\Delta_l(M)=V'_l(M(x,l)) .
\ee
We get:
\beqn
tr u [\sum_kd_{lk}\Box M(x,k)-\Delta_l(M)+
\sum_{j\ne l}c_{lj}M(x,j)]=\nonumber\\
\sum_\alpha[\sum_k(d_{lk}(m_\alpha(x,k)\Box_x+2\partial_\mu
 m_\alpha(x,k)\partial_\mu+\Box_x m_\alpha(x,k))-
\Delta_\alpha(x,k)\delta_{lk})tr u T^\alpha(x,k)+\nonumber\\
\sum_{j\ne l}c_{lj} m_\alpha(x,j)tr u T^\alpha(x,j)] .
\eeqn
We have that:
\be
tr u T^\alpha(x,l)=-i \frac{\delta tr u[\lambda]}{\delta\lambda_\alpha(x,l)} .
\ee

and
\beqn
\delta u =\int^1_0 dt e^{i t \sum_l \sum_{\alpha=1}^N
\lambda_\alpha(x,l) T^\alpha(x,l)}
i \sum_l \sum_{\alpha=1}^N \lambda_\alpha(x,l) \sum_{\beta\ne \alpha}
\nonumber\\
\frac{T^\alpha(x,l)\psi(x,l)T^\beta(x,l)+
T^\beta(x,l)\psi(x,l)
T^\alpha(x,l)}{m_\alpha(x,l)-m_\beta(x,l)}
e^{i(1-t)\sum_l \sum_{\alpha=1}^N \lambda_\alpha(x,l) T^\alpha(x,l)}
\eeqn

We compute the fermionic average to  get:
\be
\langle\delta u\bar\psi(x,l)\rangle= \langle i\sum_{\beta\ne\alpha}
\int^1_0 \frac{\lambda_\alpha(x,l)-
\lambda_\beta(x,l)}{m_\alpha(x,l)-m_\beta(x,l)} tr u[t\lambda] T^\beta(x,l)
tr u[(1-t)\lambda] T^\alpha(x,l)\rangle .
\ee

Since $T^\alpha(x,l)$ appears only once in the trace, we can express
it as a derivative of $u$:
\beqn
tr u[t\lambda] T^\beta(x,l) =\frac{-i}{t}\frac{\delta tr u[t\lambda]}
{\delta\lambda_\beta(x,l)} \nonumber\\
tr u[(1-t)\lambda] T^\alpha(x,l) =\frac{-i}{1-t}
\frac{\delta tr u[(1-t)\lambda]}{\delta\lambda_\alpha(x,l)}
\eeqn
We get the following  loop equation:
\beqn
\langle\sum_{\beta\ne\alpha}\int_0^1 dt \frac{\lambda_\alpha(x,l)-
\lambda_\beta(x,l)}{m_\alpha(x,l)-m_\beta(x,l)}\frac{1}{t(1-t)}
\frac{\delta tr u[t\lambda]}{\delta\lambda_\beta(x,l)} \frac{\delta
tr u[(1-t)\lambda]}{\delta\lambda_\alpha(x,l)}\rangle= \nonumber\\
-\langle\sum_\alpha[\sum_{j\ne l}c_{lj}m_\alpha(x,j)
\frac{\delta tr u[\lambda]}{\delta\lambda_\alpha(x,j)}+\nonumber\\
\sum_kd_{lk}(m_\alpha(x,k)\Box_x+2\partial_\mu
 m_\alpha(x,k)\partial_\mu+\Box_x m_\alpha(x,k)
-\Delta_\alpha(x,k)\delta_{lk})
\frac{\delta tr u[\lambda]}{\delta\lambda_\alpha(x,k)}
]\rangle ,
\eeqn
which is valid to any order in the $1/N$ expansion.

We can simplify the equation  if we restrict ourselves to the leading order
in $1/N$. Then we can apply Witten's factorization
property  \cite{witten}to the
loop equations to prove that:
$$
F(M)\sim \langle F(M)\rangle + O(N^{-a}) ,\ \ a>0
$$
for any $U(N)$-invariant $F$.
In particular, we obtain that:
$$
m_\alpha(x,l)\sim\langle m_\alpha(x,l)\rangle=\langle m_\alpha(0,l)\rangle .
$$
That is, in leading order of large $N$ the loop equation becomes a
closed system for $u[\lambda]$:
\beqn
\sum_{\beta\ne\alpha}\int_0^1 dt \frac{\lambda_\alpha(x,l)-
\lambda_\beta(x,l)}{m_\alpha(l)-m_\beta(l)}\frac{1}{t(1-t)}
\frac{\delta tr u[t\lambda]}{\delta\lambda_\beta(x,l)} \frac{\delta
tr u[(1-t)\lambda]}{\delta\lambda_\alpha(x,l)}= \nonumber\\
-\sum_\alpha[\sum_{j\ne l}c_{lj}m_\alpha(j)
\frac{\delta tr u[\lambda]}{\delta\lambda_\alpha(x,j)}+\nonumber\\
\sum_kd_{lk} (m_\alpha(k)\Box_x
-\Delta_\alpha(k)\delta_{lk})
\frac{\delta tr u[\lambda]}{\delta\lambda_\alpha(x,k)}
] ,
\eeqn

We could derive this closed system because we chose
to write
the Schwinger-Dyson (loop) equations in terms of $T^\alpha(x,l)$
rather than $M(x,l)$.
Also by using  the BRST Schwinger-Dyson transformation
we could avoid this difficult
change of variables in the path integral and the (highly)
non-trivial measure in the $(m_\alpha,T^\alpha)$ basis.
\vskip 1pc
The loop equations must be solved subject to the following
boundary conditions:
\beqn
tr u|_{\lambda=0}=N\nonumber\\
\frac{\delta tr u}{\delta\lambda_\alpha(x,l)}=i \nonumber\\
\frac{\delta^2 tr u}
{\delta\lambda_\alpha(x,l)\lambda_\beta(x,l)}|_{\lambda=0}=
-\delta_{\alpha\beta}
\eeqn

$u=$ constant ($\lambda$
independent) is a particular solution of the loop equations, but it
does not satisfy the boundary conditions.
We must look for a non-trivial solution.

We get an  equivalent  formulation of the loop equations
noticing  that  only  the derivative of $tr u$  appears  in  the  loop
equations. So identify:
\be
v[\lambda]_\alpha(x,l)=\frac{\delta tr u}{\delta\lambda_\alpha(x,l)} .
\ee
So that the loop equation is equivalent to the following  system  of
equations:
\beqn
\sum_{\beta\ne\alpha}\int_0^1 dt \frac{\lambda_\alpha(x,l)-
\lambda_\beta(x,l)}{m_\alpha(l)-m_\beta(l)}\frac{1}{t(1-t)}
v[t\lambda]_\beta(x,l) v[(1-t)\lambda]_\alpha(x,l)= \nonumber\\
=-\sum_\alpha[\sum_{j\ne l}c_{lj}m_\alpha(j)v[\lambda]_\alpha(x,j)+
\nonumber\\
\sum_kd_{lk}(m_\alpha(k)\Box_x-
\Delta_\alpha(k)\delta_{lk})v[\lambda]_\alpha(x,k)]\\
\frac{\delta v[\lambda]_\alpha(x,l)}{\delta \lambda_\beta(y,j)}=
\frac{\delta v[\lambda]_\beta(y,j)}{\delta \lambda_\alpha(x,l)}\\
v[\lambda]_\alpha(x,l)|_{\lambda=0}=i\\
\frac{\delta
v[\lambda]_\alpha(x,l)}{\delta\lambda_\beta(x,l)}|_{\lambda=0}=
-\delta_{\alpha\beta}
\eeqn

Notice that:
\beqn
tr M(x_1,l_1) M(x_2,l_2) ...M(x_n,l_n)=
\sum_{\alpha_1...\alpha_n} m_{\alpha_1}
...m_{\alpha_n}(l_n) tr T^{\alpha_1}(x_1,l_1)...
T^{\alpha_n}(x_n,l_n) \nonumber\\
=\frac{1}{N}\sum_{\alpha_1...\alpha_n} m_{\alpha_1}...m_{\alpha_n}
[tr T^{\alpha_1}(x_1,l_1) ..
T^{\alpha_n}(x_n,l_n)+ \rm{  all\  permutations\  of
\ indices\  \alpha}]
\eeqn

Then it is enough to know $tr u[\lambda]$
to get all correlations of the matrix
$M(x,l)$ that are symmetric under
permutations of the space-time points $(x_i,l_i)$.

A continuum version of the large-$N$ loop equation is obtained  by the usual
manipulations. That is we introduce the functions $\lambda(z,x,l)$
and $m(z,l)$
defined for $0\leq z\leq 1$ and the density of eigenvalues $\rho_l$, and make
the
following identifications:
\beqn
\lambda_\alpha(x,l)=\lambda(\frac{\alpha}{N},x,l)\\
m_\alpha(l)=\sqrt{N} m(\frac{\alpha}{N},l)\\
\rho_l(m_l)=\frac{dx_l}{dm_l}\\
\int dm_l \rho_l(m_l)=1 .
\eeqn

The large-$N$ loop equation becomes:
\beqn
P\int dm_1 dm_2\rho_l(m_1)\rho_l(m_2)\int_0^1 dt\frac{\lambda(m_1,y,l)-
\lambda(m_2,y,l)}{m_1-m_2}\frac{1}{t(1-t)}
\times\nonumber\\
\frac{\delta tr u[t\lambda]}{\delta\lambda(m_2,y,l)}
\frac{\delta tr u[(1-t)\lambda]}{\delta\lambda(m_1,y,l)}=
-\sum_{j\ne l} c_{lj}\int dm \rho_j(m) m
\frac{tr u[\lambda]}{\delta\lambda(m,y,j)}\nonumber\\
-\int dm\rho_l(m)(m\Box_y-\Delta_l(m))
\frac{tr u[\lambda]}{\delta\lambda(m,y,l)} ;
\eeqn
$P$ stands for the principal value of the integral over $m_i$.

\section{The Zero dimensional case}

In this section we discuss the zero dimensional case. The equations are:

\beqn
\sum_{\beta\ne\alpha}\int_0^1 dt \frac{\lambda_\alpha(l)-
\lambda_\beta(l)}{m_\alpha(l)-m_\beta(l)}\frac{1}{t(1-t)}
\frac{\delta tr u[t\lambda]}{\delta\lambda_\beta(l)} \frac{\delta
tr u[(1-t)\lambda]}{\delta\lambda_\alpha(l)}= \nonumber\\
-\sum_\alpha[\sum_{j\ne l}c_{lj}m_\alpha(j)
\frac{\delta tr u[\lambda]}{\delta\lambda_\alpha(j)}-\Delta_\alpha(l)
\frac{\delta tr u[\lambda]}{\delta\lambda_\alpha(l)}] .
\eeqn
We want to check that we reproduce the equations of \cite{retamal} for
$n=2$. In this case we have:
\beqn
u[\lambda]=1+i\sum_{\alpha,l}\lambda_\alpha(l)+\frac{i^2}{2}[\sum_\alpha(
\lambda_\alpha(1)^2+\lambda_\alpha(2)^2+\nonumber\\
2\sum_{\alpha,\beta}\lambda_\alpha(1)
\lambda_\beta(2)tr(T_\alpha(1)T_\alpha(2))]+
\frac{i^3}{6}[\sum_\alpha\lambda_\alpha(1)^3+\lambda_\alpha(2)^3+\nonumber\\
3\sum_{\alpha,\beta}(\lambda_\alpha(1)^2\lambda_\beta(2)+\lambda_\alpha(1)
\lambda_\beta(2)^2) tr(T_\alpha(1)T_\beta(2))+o(\lambda^4)
\eeqn
Inserting this into the loop equations
we readily get the system of \cite{retamal}.
In the same way we can get the equations
for the higher correlators of $T_\alpha(1),
T_\beta(2)$.

It is, of course, a very interesting
problem to see if the methods of \cite{alfaro1}
can be extended to the n-matrix chain.

\section{Discussion and Open Problems}

We have been able to derive the loop equations for the d-dimensional n-matrix
model. They are a consequence of the Schwinger-Dyson equations satisfied
by the Green functions of the model. Due to the factorization property
of U(N)-invariant operators in large $N$ the loop equations form a closed
set that may be the starting point for a non-perturbative description of
the system. In particular our result applies to the q-state Potts matter
fields coupled to two dimensional gravity that have been previously proposed
to describe the non-critical string with $c>1$.

There are several interesting open problems.
It is known that the loop equations
for $d=0$ models can be realized as
Virasoro\cite{verlinde} and $W_3$ \cite{narain}
constraints on the free energy
of the system. We wonder whether this is true for the present case too.
In addition the loop equations could be used as the starting point for
a different approximation to the Physics of the model, perhaps on the lines
suggested by the solution of the Two matrix model in \cite{alfaro1}.

\section*{Acknowledgements}

This work has been partially supported by Fondecyt \# 1930566.

\end{document}